\documentclass[reprint,superscriptaddress,amsmath,amsfonts,amssymb,aps,prl,showkeys]{revtex4-1}
\usepackage{bm}
\usepackage{color}
\usepackage{graphicx}

\usepackage{hyperref}
\hypersetup{
  breaklinks = {true},
  citecolor = {blue},
  colorlinks = {true},
  linkcolor = {red},
}

\begin{document}

\title{Canted Spin Texture and Quantum Spin Hall Effect in WTe$_2$}

 \author{Jose H. Garcia}
\affiliation{Catalan Institute of Nanoscience and Nanotechnology (ICN2), CSIC and BIST, Campus UAB, Bellaterra, 08193 Barcelona, Spain}
\author{Marc Vila}
\affiliation{Catalan Institute of Nanoscience and Nanotechnology (ICN2), CSIC and BIST, Campus UAB, Bellaterra, 08193 Barcelona, Spain}
\affiliation{Department of Physics, Universitat Aut\`onoma de Barcelona, Campus UAB, Bellaterra, 08193 Barcelona, Spain}
\author{Chuang-Han Hsu}
\affiliation{Department of Electrical and Computer Engineering, National University of Singapore, Singapore 117576, Singapore }
\author{Xavier Waintal}
\affiliation{Univ. Grenoble Alpes, CEA, IRIG-PHELIQS, 38000 Grenoble, France}
\author{Vitor M. Pereira}
\affiliation{Centre for Advanced 2D Materials and Graphene Research Centre, National University of Singapore, Singapore 117546, Singapore}
\author{Stephan Roche}
\affiliation{Catalan Institute of Nanoscience and Nanotechnology (ICN2), CSIC and BIST,
Campus UAB, Bellaterra, 08193 Barcelona, Spain}
\affiliation{ICREA--Instituci\'o Catalana de Recerca i Estudis Avan\c{c}ats, 08010 Barcelona, Spain}

\date{\today}

\begin{abstract}
We report an unconventional quantum spin Hall  phase in the monolayer T$_\text{d}$-WTe$_2$, which exhibits hitherto unknown features in other topological materials. The low-symmetry of the structure induces a canted spin texture in the $yz$ plane, which dictates the spin polarization of topologically protected boundary states. Additionally, the spin Hall conductivity gets quantized ($2e^2/h$) with a spin quantization axis parallel to the canting direction.
These findings are based on large-scale quantum simulations of the spin Hall conductivity tensor and nonlocal resistances in multi-probe geometries using a realistic tight-binding model elaborated from first-principle methods.
The observation of this canted quantum spin Hall effect, related to the formation of topological edge states with nontrivial spin polarization, demands for specific experimental design and suggests interesting alternatives for manipulating spin information in topological materials.
\end{abstract}

\maketitle

{\it Introduction.}
The prediction of the quantum spin Hall (QSH) insulator state \cite{Kane2005, Kane2005QSH, Sheng2005,Sheng2006,Fu2007} and its connection with topological states in strong spin-orbit coupling materials \cite{Bernevig1757, Bernevig2006} sparked an exciting playground for fundamental studies \cite{HasanRMP2010, Sinova2015, Yang2016a}. The subsequent demonstration of the existence of topological insulators \cite{Bernevig1757, Konig766, Hsieh2008, Jiang2009, Zhang2009, Ren2016} then opened a myriad of technological possibilities, since topologically protected states are predicted to carry spin information over unprecedented distances due to a strong resilience to disorder, as long as time-reversal symmetry is preserved \cite{Sheng2006, Bernevig2006, Ezawa2014, Dayi2016}. But, to date, a QSH effect at room-temperature has not yet been experimentally achieved \cite{Konig2007, HasanRMP2010, Ortmann2015, Tang2017, Fei2017, Jia2017, Chen2018, Wu2018, Shi2019, Reis2017}. 
In this context, two-dimensional transition metal dichalcogenides (TMDs) in their $1{\rm T}^\prime$ $(P2_{1}/m)$ and $1{\rm T}_\text{d}$ $(Pmn2_{1}$) structural phases are seen as ideal platforms to engineer a resilient QSH regime as well as novel electronic devices driven by an electric-field-tunable topological phase transition \cite{Qian2014}. Recent signatures of the QSH effect up to $\sim$\,100\,K in monolayer WTe$_2$ \cite{Wu2018} are very encouraging. However, the lack of a robust Hall conductance quantization insensitive to the device characteristics\,---\,hallmark of topological physics in the quantum Hall regime \cite{PhysRevLett.45.494}\,---\,demands in-depth scrutiny of the fundamentals of spin transport in both the topologically trivial and nontrivial regimes, as well as an assessment of any underlying fundamental limitations \cite{Peng2017, Fei2017, Ok2019, Vayrynen2019, Novelli2019}.  

On the other hand, the traditional spin Hall effect, driven by spin-dependent impurity scattering, is usually associated with spin polarization pointing perpendicular to the conducting plane \cite{Hirsch1999,Sinova2015}. Some models of 2D QSH systems, such as the Kane-Mele-Haldane Hamiltonian  in the absence of Rashba spin-orbit coupling (SOC) terms, are characterized by helical edge states whose spin are also perpendicularly polarized \cite{Kane2005, Sheng2005}. However, an out-of-plane spin polarization is not an inherent property of the intrinsic spin Hall effect (quantized or not), but rather a consequence of the underlying symmetries of the crystal. 
As a matter of fact, different experimental groups recently measured spin Hall conductivities  associated with both in- and out-of-plane spin polarization components in few-layer 1T$_\text{d}$ and 1T$^\prime$-MoTe$_2$, both of similar magnitude, illustrating peculiar aspects of bulk spin transport in these materials \cite{Safeer2019, Song2020,Zhao2020,Zhao2020b,Seemann2015, Song2020}. However, to date, little is known about the imprint of the inherently low symmetry of this class of TMDs in the QSH regime. Correlations and substrate effects were found to induce localization of edge modes \cite{Ok2019}, but the impact of low-symmetries and multiple spin Hall components in the QSH remains to be determined. 

In this Letter, we show that the low symmetry phase  ($1{\rm T}_{\rm d}$) of the WTe$_2$ monolayer leads to an unconventional QSH effect, in which the topological edges states exhibit a canted spin polarization in the $yz$ plane. This differs from the conventional $z$-polarized feature frequently discussed for canonical models of QSH systems. Moreover, the spin Hall conductivities becomes quantized in contrast with other QSH topological insulators \cite{Kane2005QSH}. 
These results emerge from complementary calculations of the spin Hall conductivity tensor combined with simulations of nonlocal transport in realistic multi-probe geometries, with and without disorder. The calculations hinge upon an effective 4-band tight-binding model that, beyond symmetry, reproduces the essential features of the low-energy band and spin structures of this material. The results further reveal that the spintronic potential of WTe$_2$ is unique even when doped away from the QSH insulator regime, as it displays a peculiar spin texture, defined as a constant spin polarization throughout the entire Fermi contour \cite{Schliemann2003,Bernevig2006,Schliemann2017}. 
In addition to numerical calculations with millions of orbitals, we provide analytical connections between the canted spin quantization axis of the topological states and the spin texture of the bulk bands, induced by the SOC parameters.
.

{\it Model and methodology.}
We derived a generic DFT-based 4-band tight-binding Hamiltonian on a rectangular lattice, which is applicable to 1T$^\prime$ and 1T$_\text{d}$ TMDs as discussed in detail elsewhere \cite{Unpublished}. The spinful model describes the two lowest energy bands belonging to the irreducible representations $A_g$ (conduction, mostly of metal $d$ orbital content) and $B_u$ (valence, of $p$ orbital content) of the point group $C_{\text{2h}}$. This pair of bands is ``inverted'' in  WTe$_2$ at the $\Gamma$ point, rendering the ground state a QSH topological insulator. The model is similar to symmetry-based models used in recent work \cite{Xu2018, Shi2019prb,Ok2019}. When expanded near the $\Gamma$ point, the Hamiltonian ${\cal H} = {\cal H}_0 + {\cal H}_\text{soc}$ has the following $\bm{k}\,{\cdot}\,\bm{p}$ representation:
\begin{equation}
{\cal H}_0 \simeq (k_x^2+k_y^2) (m_p\tau_0+m_d\tau_z) +\beta k_y  \tau_y + \delta\tau_z + \eta \tau_x,
\label{eq:H0}
\end{equation}
with $\tau_i$ $(i=x,y,z)$ the $2\times 2$ Pauli matrices in the space spanned by the two orbitals. The parameters $m_p=-0.1050$ eV and $m_d=-0.5449$ eV are related to the effective masses of the valence and conduction bands, $\delta=0.4248$ eV describes the degree of band inversion at $\Gamma$, $\beta=0.4494$ eV models the $x$--$y$ crystalline anisotropy ($\hat{\bm{x}}{\,\parallel}\,\bm{a}$ crystal axis or zig-zag direction), and $\eta$ breaks inversion symmetry to describe either the $1{\rm T}^\prime$ ($\eta$ = 0) or $1{\rm T}_\text{d}$($\eta$=0.0017 eV) phase. 
At the $\bm{k}\,{\cdot}\,\bm{p}$ level, the SOC is given by 
\begin{equation}
{\cal H}_{\rm soc} \simeq (\Lambda_x k_y \sigma_x +\Lambda_y k_x \sigma_y +\Lambda_z k_x \sigma_z  )\otimes \tau_x,
\label{eq:Hsoc}
\end{equation}
where $(\Lambda_x,\Lambda_y,\Lambda_z) = (0.0591,0.0777,-0.1159)$ eV and the Pauli matrices $\sigma_i$ $( i=x,y,z)$ are defined in the spin space. The parameters were determined by fitting the band structure and spin textures to reproduce DFT calculations as described in Reference \onlinecite{Unpublished}. 

\begin{figure}
\includegraphics[width =0.48\textwidth]{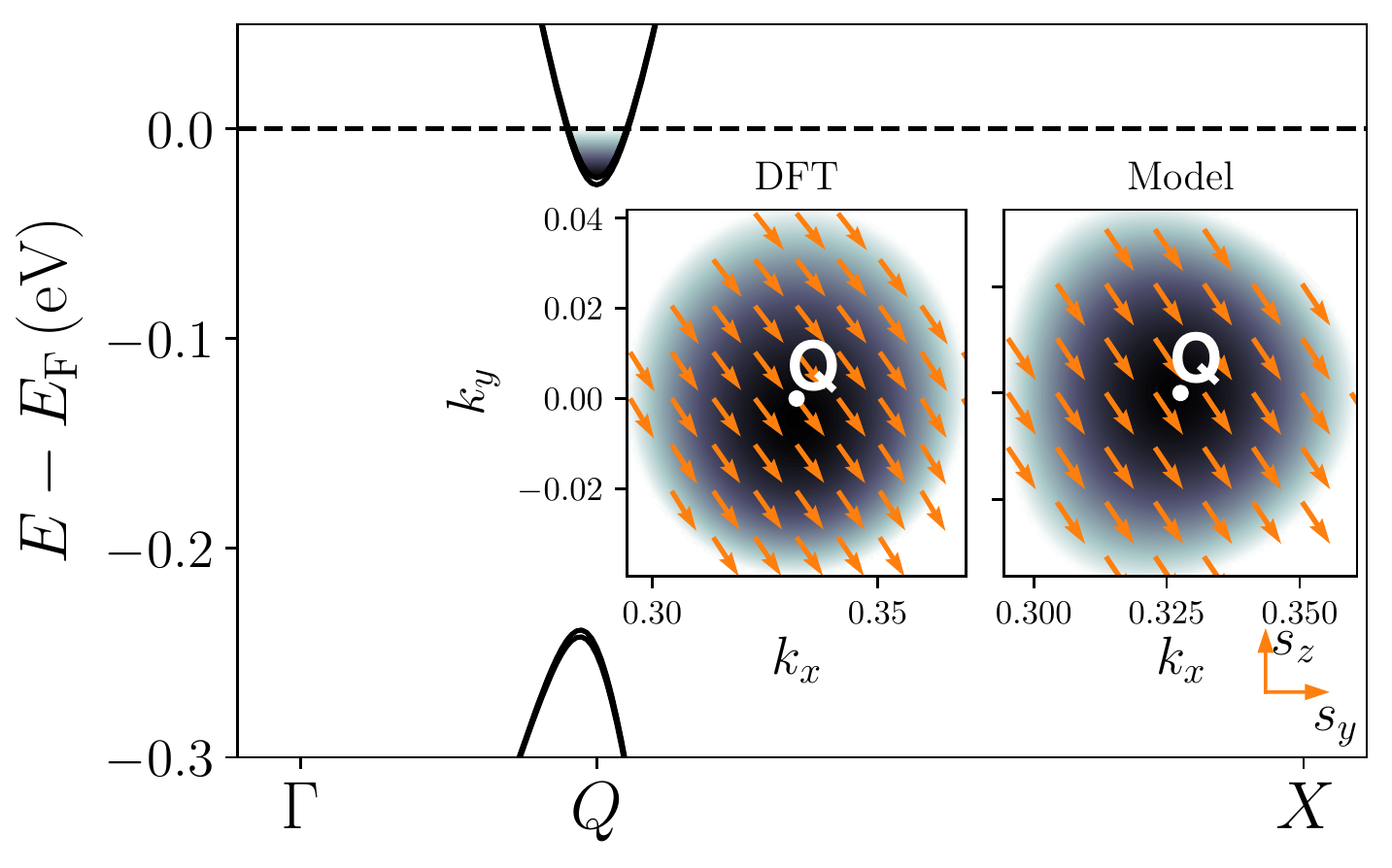}
\caption{Band structure of WTe$_2$ around the charge pockets formed by the band inversion at $\Gamma$. The conduction band minimum is at the $k$-point $\bm{Q}=(0.332,0.0)$ with energy $\varepsilon_0 \approx -27$ meV. The inset compares the spin textures computed from DFT and the effective model; the color represents the energy with respect to the Fermi level and the arrows the spin orientation in the $yz$ plane (the spin projection along $x$ is negligible). The white dots indicates the position of the $Q$-point. All $k$-points are in units of $\pi/a$ with $a$ the lattice constant along the zig-zag direction.  
}
\label{fig_F1}
\end{figure}

Fig.~\ref{fig_F1} shows a close-up of the model-generated band structure near the Fermi level. The underlying DFT calculation is based on the PBE + HSE functional \cite{Heyd2003}, which places the Fermi level ($E_F$) near the bottom of the conduction band. The effective model describes accurately the conduction band and energy gap. Each band features two charge pockets symmetrically located away from $\Gamma$, with minima ($\varepsilon_0\approx -27$ meV) at the point labeled $Q$ and its time-reversal counterpart (not shown). Though present, the spin-orbit band splitting is too small and barely discernible at the scale shown. 
In the inset, we compare the spin textures at $E_F$ for one of the $Q$-centered electron pockets obtained by DFT at $(0.332,0.0)$ with that arising from the model at $(0.327,0.0)$, both in units of $\pi/a$. The spin orientations in the $yz$ plane are represented by the orange arrows (despite not strictly zero, the $x$ component is omitted for clarity, as it was found comparatively much smaller in magnitude). In addition to the obvious agreement, it is noteworthy that the spin texture is constant to a very good approximation. WTe$_2$ is hence a case with a naturally present persistent spin texture which is invariant upon changing $E_F$ within the range of energies shown. The spins cant at an angle $\theta\approx -56^\circ$ with respect to $y$. 

\begin{figure}
\includegraphics[width =0.5\textwidth]{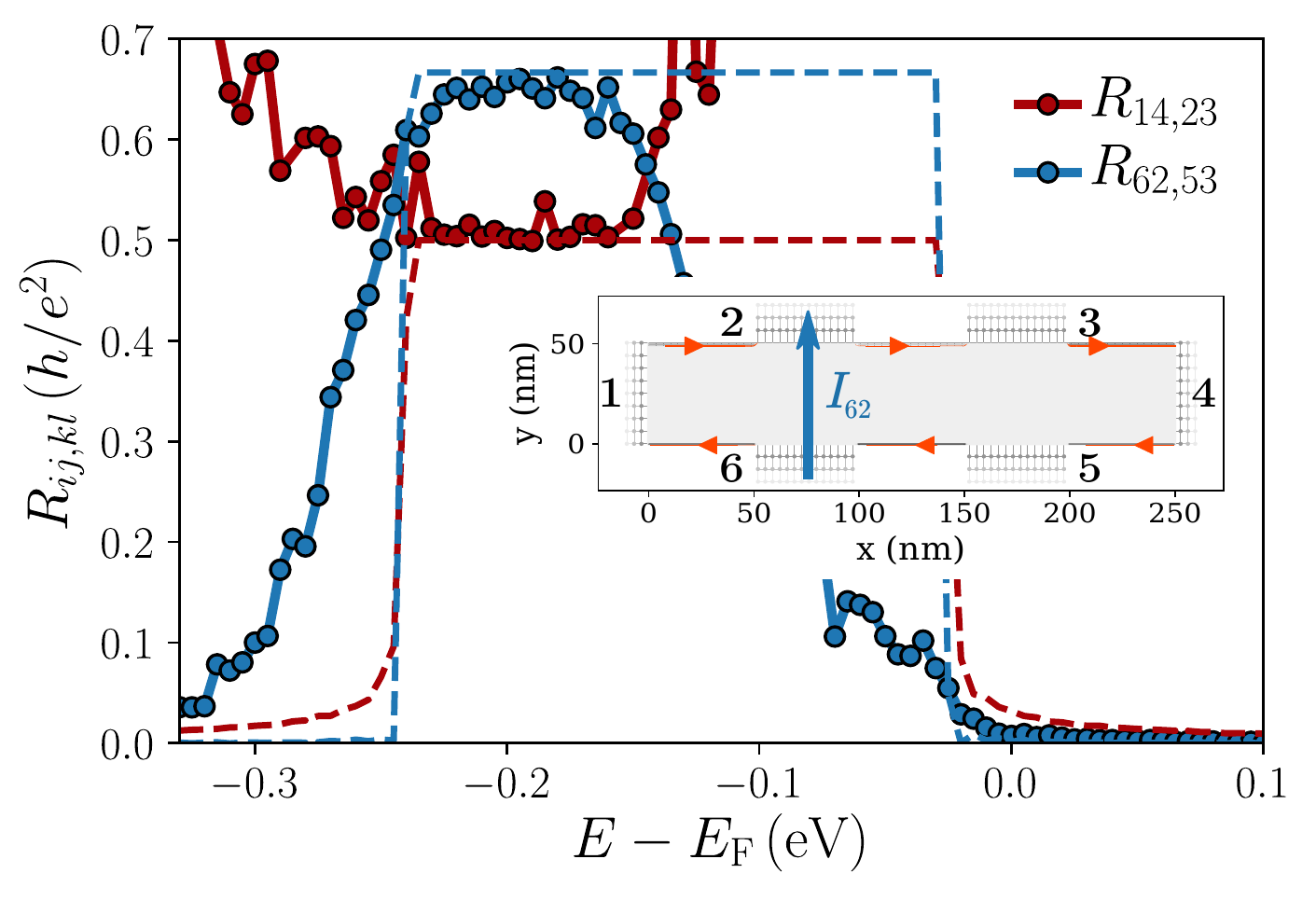}
\caption{Nonlocal resistances $R_{ij,kl} = (V_k-V_l)/I_{ij}$ calculated in the 6-terminal Hall-bar device shown in the inset. The two plateau values $2h/3e^2$ and $h/2e^2$ seen here unequivocally attribute the nonlocal signal to QSH edge states \cite{Roth2009}. Solid (dashed) lines correspond to simulations with (without) Anderson disorder (with strength $U=2$ eV). In the inset, the solid (lattice) regions delineate the device (leads). The device is defined on a rectangular lattice (parameters $a_x = 3.4607$ \AA{} and $a_y = 6.3066$ \AA{}). The device width, inter-lead separations, and lead widths are all 50 nm. The small horizontal arrows along the top and bottom edges mark the direction of the local, bond-projected spin current density $\bm{J}^{z'}_s$ arising as a response of driving charge current from lead 6 to lead 2.}

\label{fig_F3}\end{figure}

We next explore the nature of spin transport as $E_F$ is varied across the band gap by computing the spin Hall conductivity tensor ($\sigma_{ij}^{\alpha}, \alpha=x,y,z$) using the Kubo-Bastin formula implemented for the tight-binding model \cite{Bastin1971, cresti2016rnc}:
\begin{align}\label{eq_Kubo}
\sigma_{ij}^{\alpha}=-2\hbar\Omega  \int_{-\infty}^{E_{\rm F}} dE\, \text{Im}\left( \text{Tr}\left[  \delta(E-{\cal H})J_{s,i}^{\alpha}\frac{dG^+}{dE}J_{j} \right]\right),
\end{align}
where $\Omega$ is the area; $J_{s,i}^\alpha \equiv \{J_{i},\sigma_\alpha\}/2\,$ is the $i$-th component of the spin current density operator, with $\alpha=x,y,z$ denoting the spin polarization direction; $J_{j} \equiv (ie/\Omega\hbar) [ {\cal H} , R_{j}]$ is the $j$-th component of the current density operator, with $e$ the electron charge and $R_{j}$ the position operator \cite{Fan2019}. The spectral operators $\delta(E-{\cal H})$ and $G^+\equiv 1/(E-{\cal H}+i0^+)$ are the Dirac delta and the retarded Green's function, respectively. We numerically computed the Kubo-Bastin formula by using the kernel polynomial method \cite{Garcia2015, cresti2016rnc, Garcia2018, Fan2019} with 2000 Chebyshev expansion moments (which is equivalent to a broadening of 5\,meV). These calculations were carried out on a system containing $4\times1000\times1000$ orbitals. In addition, we simulated multi-terminal nonlocal transport within the Landauer-B\"{u}ttiker framework as implemented in the Kwant package \cite{Groth2014, Vila2020}, using the six-terminal device geometry shown in the inset of Fig.~\ref{fig_F3}.

\begin{figure}
\includegraphics[width =0.5\textwidth]{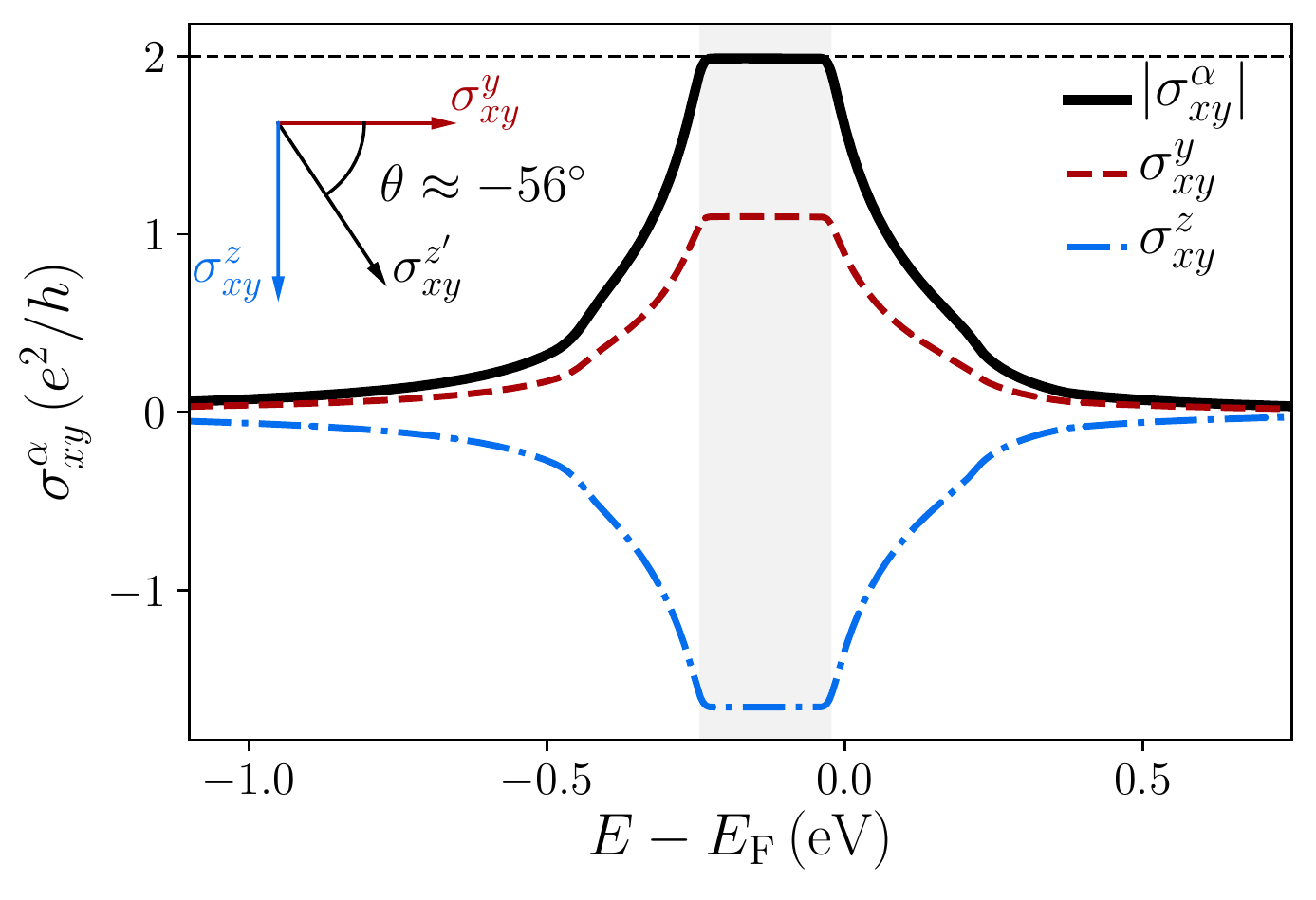}
\caption{Spin Hall conductivities $\sigma_{xy}^y$ and $\sigma_{xy}^z$. The solid line shows the norm of $|\sigma_{xy}^\alpha| \equiv \sqrt{(\sigma_{xy}^y)^2 + (\sigma_{xy}^z)^2}$. Inset: orientation of the spin of the helical edge states. The calculations were done considering a broadening of 5 meV on a system with $1000\times1000\times4$ orbitals.}
\label{fig_F2}
\end{figure}

{\it Spin Hall conductivity.}
Fig.~\ref{fig_F2} shows the non-zero components of the transverse spin Hall conductivity tensor, $\sigma^{\alpha}_{xy}$, $\alpha\in\{y,z\}$, as $E_F$ is varied near and within the band gap. Although both $\sigma^{z}_{xy}$ and $\sigma^{y}_{xy}$ display a plateau in the gap region, their values are $-1.65\,e^2/h$ and $1.1\,e^2/h$, respectively. This is intriguing since usually, within a topological gap, quantized spin Hall conductivities are integer multiple of the conductance quantum, reflecting the existence of a definite (integer) number of helical edge channels. 

However we note that, by definition, each component $\alpha$ of $\sigma^\alpha_{xy}$ provides only a measure of the \emph{projection} of the spin onto the Cartesian direction $\alpha$, because $\sigma^\alpha_{ij} \propto J^{\alpha}_{s,i} / J_j$ where $\bm{J}^{\alpha}_{s}$ is the spin current density carrying spins polarized parallel to $\alpha$ in response to a driving charge current $\bm{J}$. But the choice of Cartesian directions is arbitrary\,---\,in fact, the results in Fig.~\ref{fig_F2} show that a Cartesian system fixed by the orthorhombic axes of the crystal obscures the adequate spin quantization axis in this problem. 
This is readily confirmed by the fact that, in the gap, $|\sigma_{xy}^\alpha| \equiv \sqrt{ (\sigma_{xy}^y)^2 + (\sigma_{xy}^z)^2 }$ is indeed quantized at $2 e^2/h$ (Fig.~\ref{fig_F2}, solid curve), where the factor of 2 reflects the existence of two counter-propagating modes per edge.
This shows that the interdependence among the magnitudes of the spin Hall conductivities components seen in Fig.~\ref{fig_F2} stems from a fundamental common origin, namely the presence of \emph{spin-canted topological edge states} which sustain a QSH effect in WTe$_2$.
From the values of each plateau, we determine that the spin quantization axis is canted at $\arctan{(\sigma_{xy}^z/\sigma_{xy}^y)} = -56^\circ$ with respect to the $y$ axis. Notably, this angle matches perfectly with the orientation of the persistent spin texture near the bottom of the conduction band, shown earlier in the inset of Fig.~\ref{fig_F1}. 

To elucidate the origin of this behavior more explicitly, we unitarily transform the Hamiltonian $\cal H$ with a rotation in spin space about $\hat{\bm{x}}$, which is effected by the matrix $U(\theta)\equiv\cos[\,(2\theta-\pi)/4\,]\sigma_0-i\sin[\,(2\theta-\pi)/4\,]\sigma_x\,$, where $\theta\,{\equiv}\,\arctan{(\Lambda_z/\Lambda_y)}\,{\approx}\,{-}\,56^\circ$ is an angle defined by the SOC parameters in Eq.~\eqref{eq:Hsoc}. While ${\cal H}_0$ is invariant under this operation, the SOC term transforms into 
\begin{equation}
  {\cal H}_{\rm SOC}' \equiv U^\dagger(\theta){\cal H}U(\theta) 
  = \Lambda_x k_y \sigma_x + \Lambda_r k_x \sigma_{z}'\tau_x,
  \label{eq:Hsoc2}
\end{equation}
with $\Lambda_r \equiv \sqrt{\Lambda_z^2+\Lambda_y^2}$ and $\sigma_{z'} \equiv U^\dagger(\theta)\sigma_z U(\theta)$. We now note that $\Lambda_x$ is numerically smaller than $\Lambda_r$ in WTe$_2$ and, more importantly, $|k_y| \ll |k_x|$ near the bottom of the $Q$-centered electronic pockets. The combined effect is that, over the range of energies shown in Fig.~\ref{fig_F1}, the first term in Eq.~\eqref{eq:Hsoc2} is two orders of magnitude smaller than the second and thus \emph{negligible in practice}. Consequently, $[{\cal H}',\sigma_{z'}] \approx 0$ so that spin is preserved along the canted $z'$ direction to a very good approximation, which has two physical consequences: (i) when $E_F$ lies in the conduction band, the carriers have a persistent spin texture directed along $z'$ over the entire Fermi contour; (ii) the canting angle is preserved in the QSH regime (when $E_F$ lies in the gap), which supports the quantization of the spin Hall conductivities and defines a canted QSH effect. 

{\it Chiral transport of spin at the edges.}
The topological nature of the electronic states can be unequivocally confirmed by probing nonlocal resistances, $R_\text{nl}$, in a Hall-bar geometry under different bias conditions: If the nonlocal signal is due only to helical edge states, $R_\text{nl}$ should display plateau values uniquely determined by the specific combination of contacts chosen for current injection and nonlocal voltage detection \cite{Roth2009}. We employed our effective tight-binding model to compute the nonlocal resistance using the device geometry illustrated in Fig.~\ref{fig_F3}(inset). To obtain $R_\text{nl}$, we first calculate the Landauer-B\"uttiker transmission probabilities between each pair of leads and build the conductance matrix $G_{ij}$  \cite{Datta1997,Vila2020} that satisfies the linear system of equations $I_i=\sum_j G_{ij} V_j$, where $I_i$ and $V_j$ describe the current and voltage at each lead. We then require the current to flow from lead $i$ to $j$ by setting $I_i=-I_j$ and $I_k=0, k\neq i,j$, and calculate the resulting voltages $V_j$. The nonlocal resistances are defined as $R_{ij,kl} \equiv (V_k-V_l)/I_{ij}$, i.e., current flows from lead $i$ to $j$ and voltage is measured between leads $k$ and $l$. Furthermore, to test the robustness of the nonlocal signal, we included (non-magnetic) Anderson disorder in the tight-binding Hamiltonian, diagonal in both orbital and spin spaces, whereby a uniformly distributed random energy $U_r$ is added at each lattice site, with $U_r\in[-U/2,U/2]$.

The results are plotted in Fig \ref{fig_F3}, where solid (dashed) lines show $R_\text{nl}$ for a system with (without) disorder. Each curve represents a different calculation of $R_\text{nl}$, that is, a different choice of current paths and probes used to calculate $R_{ij,kl}$. The quantized values obtained at the plateaus precisely correspond to those expected in the QSH state for the chosen injection and detection contacts \cite{Roth2009}. The fact that different choices of electrical contacts yield distinct\,---\,yet precisely defined\,---\,plateau values stems from the equilibration condition of the chemical potential at the leads \cite{Roth2009}; therefore, the chosen voltage probes and the current path uniquely determine the value of $R_{ij,kl}$.  Note however that such nonlocal setup is unable to discern the $y$ and $z$ projections of the spin in the edge states, for that, one may need to use magnetic electrodes. 

We also computed the bond-projected \emph{spin} currents \cite{Groth2014} for spins polarized along the (rotated) ${z'}$ and ${y'}$ directions, i.e., $\bm{J}^{z'}_s$ and $\bm{J}^{y'}_s$. The former is shown in the inset of Fig.~\ref{fig_F3} as horizontal arrows at the top/bottom edges, evidencing the fingerprint of helical transport in the QSH regime. In contrast, $\bm{J}^{y'}_s$ was found to be negligible, which is consistent with the form of ${\cal H}_\text{soc}'$ in Eq.~\eqref{eq:Hsoc2}. Finally, we also observe a strong resilience of the plateaus to the presence of nonmagnetic disorder, see Fig \ref{fig_F3}, consistent with time-reversal topologically protected states ($U=2$ eV much larger than any other energy scale of the Hamiltonian). These nonlocal results clearly establish that the canted QSH effect, inferred above from a bulk Kubo calculation, is characterized by robust helical spin transport at the edges, a fact fully consistent with the bulk-boundary correspondence \cite{Hatsugai1993}. 

{\it Conclusion.}
We performed quantum transport simulations which allowed an in-depth study of the nature of spin transport in monolayers of WTe$_2$, with emphasis on the QSH regime. Calculations of spin Hall conductivities and nonlocal resistances in multi-probe configurations revealed a so-far-unique QSH effect regime defined by a canted spin quantization axis, fixed by SOC characteristics. The oblique spin polarization of topological edge states in the QSH regime is related to a persistent spin texture for Fermi level placed in the conduction band.
Our findings call for a careful analysis of QSH effect measurements, whose interpretation usually ignores the possibility of multiple non-zero components of the spin Hall conductivity tensor\,---\,as a result, non-integer quantization might be erroneously inferred by improper measurement design. Also, such non-integer QSH plateaus were theoretically discussed for square and hexagonal lattices \cite{Matusalem2019}, suggesting the possible existence of a canted QSH effect in other systems as well. A combination of measurements with applied magnetic field along different directions, or nonlocal measurements with magnetic contacts, could disentangle the different contributions of such peculiar topological spin dynamics. 
The low-symmetry phases of TMDs may thus provide fascinating avenues to design new topological nanodevices for spin transport beyond the current paradigm of QSH effect with $z$-polarized spins. 

\begin{acknowledgments}
M.V. acknowledges support from ``La Caixa'' Foundation and the Centre for Advanced 2D Materials at the National University of Singapore for its hospitality. X. W. acknowledges the ANR Flagera GRANSPORT funding. ICN2 authors were supported by the European Union Horizon 2020 research and innovation programme under Grant Agreement No. 881603 (Graphene Flagship) and  No.  824140 (TOCHA, H2020-FETPROACT-01-2018). ICN2 is funded by the CERCA Programme/Generalitat de Catalunya, and is supported by the Severo Ochoa program from Spanish MINECO (Grant No. SEV-2017-0706). V.M.P. acknowledges the support of the National Research Foundation (Singapore) under its Medium-Sized Centre Programme (R-723-000-001-281).
\end{acknowledgments}

\bibliography{bibmoteQSHE}

\end{document}